\documentclass[12pt]{article}

\usepackage[left=2cm,top=2cm,right=2cm,bottom=2cm]{geometry}
\usepackage[english]{babel}
\addto{\captionsenglish}{}
\usepackage{graphicx}
\usepackage[square]{natbib}
\usepackage{amsmath}
\usepackage{cite}
\usepackage{hyperref}
\usepackage[title]{appendix}
\usepackage[dvipsnames,svgnames,x11names,hyperref]{xcolor}
\hypersetup{
    colorlinks=false,
    linkcolor=black,
    filecolor=black,      
    urlcolor=blue,
}
\usepackage[nameinlink,noabbrev,capitalise]{cleveref}
\usepackage{graphics}

\usepackage{amssymb,amsmath,latexsym,amsthm,amsfonts}
\usepackage{enumerate}
\usepackage{bbm}
\usepackage{cancel}
\usepackage{caption}
\usepackage{dsfont}
\usepackage{pbox}
\usepackage{booktabs,dcolumn}
\usepackage{url}
\usepackage{siunitx}
\usepackage{makecell}
\usepackage{adjustbox}
\usepackage{subfigure}
\usepackage{enumerate}
\usepackage[utf8]{inputenc}
\usepackage{longtable}
\usepackage{graphicx}    
\usepackage{mathrsfs} 
\usepackage{multicol}
\usepackage{bm}
\usepackage{appendix}
\usepackage{float}
\usepackage{titlesec}

\titleformat*{\section}{\LARGE\bfseries}
\titleformat*{\subsection}{\Large\bfseries}
\titleformat*{\subsubsection}{\bfseries}
\titleformat*{\paragraph}{\large\bfseries}
\titleformat*{\subparagraph}{\large\bfseries}
\usepackage{chngcntr}

\makeatletter
\def\@seccntformat#1{\@ifundefined{#1@cntformat}%
   {\csname the#1\endcsname\quad}%      default
   {\csname #1@cntformat\endcsname}%    enable individual control
}
\makeatother

\title{Low-rank representation of head impact kinematics: \\ A data-driven emulator}
\date{}

\author{Patricio Arrue$^{1}$, Nima Toosizadeh$^{1,2,3}$, Hessam Babaee$^{4}$, Kaveh Laksari$^{1,5}$\footnote{Correspondence: klaksari@arizona.edu} \\ \\ $^{1}$Dept. of Biomedical Engineering, University of Arizona\\$^{2}$Arizona Center on Aging (ACOA), Dept. of Medicine, University of Arizona\\$^{3}$Division of Geriatrics, General Internal Medicine and Palliative Medicine,\\ Dept. of Medicine, University of Arizona\\$^{4}$Dept. of  Mechanical Engineering and Material Sciences, University of Pittsburgh\\$^{5}$Dept. of Aerospace and Mechanical Engineering, University of Arizona}

\begin{document}
\maketitle

% ##############################################
% ##############################################

% ##############################################
% \vspace{2mm}

\section*{Abstract}
{\footnotesize

Head motion induced by impacts has been deemed as one of the most important measures in brain injury prediction, given that the vast majority of brain injury metrics use head kinematics as input. Recently, researchers have focused on using fast approaches, such as machine learning, to approximate brain deformation in real time for early brain injury diagnosis. 
However, training such  models requires large number of kinematic measurements, and therefore data augmentation is required given the limited on-field measured data available. 
In this study we present a principal component analysis-based method that emulates an empirical low-rank substitution for head impact kinematics, while requiring low computational cost. 
In characterizing our existing data set of 537 head impacts, each consisting of 6 degrees of freedom measurements, we found that only a few modes, e.g. 15 in the case of angular velocity, is sufficient for accurate reconstruction of the entire data set. 
Furthermore, these modes are predominantly low frequency since over 70\%  of the angular velocity response can be captured by modes that have frequencies under 40Hz. We compared our proposed method against existing impact parametrization methods and showed significantly better performance in injury prediction using a range of kinematic-based metrics -- such as head injury criterion (HIC), rotational injury criterion (RIC) and brain injury metric (BrIC) -- and brain tissue deformation-based metrics -- such as brain angle metric (BAM), maximum principal strain (MPS) and axonal fiber strains (FS). In all cases, our approach reproduced injury metrics similar to the ground truth measurements with no significant difference, whereas the existing methods obtained significantly different ($p<0.01$) values as well as substantial differences in injury classification sensitivity and specificity. 
   
This emulator will enable us to provide the necessary data augmentation to build a head impact kinematic data set of any size. The emulator and corresponding examples are available at our website \footnote{ \href{https://uweb.engr.arizona.edu/~klaksari/Resources.html}{https://uweb.engr.arizona.edu/$\sim$klaksari/Resources.html}}.

}

\section*{Introduction}
Traumatic brain injury (TBI) is one of the most debilitating health problems in our society today, with nearly two million new cases in the US every year \citep{Taylor2017}. The majority of these cases are considered mild, also known as concussion \citep{DefenseandVeteransBrainInjuryCenter2018}. The substantial increase in reported concussions in contact sports \citep{Selassie2013}, together with the recent findings of increased long-term pathological changes \citep{Dekosky2013}, has sparked a public discussion and raised awareness about TBI. An important requirement is an accurate and objective diagnosis of concussions, which in turn could inform better protective equipment design and safer activities \citep{Kuo2017a,Kurt2017,Siegkas2019,Manoogian2006,Wu2014a}. 
\\

Head motion kinematics, including the rate, frequency and direction of head's movement during collision, has been deemed as one of the most consequential metric in predicting brain injury. Historically, kinematic-based metrics such as head injury criterion (HIC) \citep{NTSA1972}, rotational injury criterion (RIC) \citep{Kimpara2012a}, and brain injury criterion (BrIC) \citep{Takhounts2013} have been used to detect injury. These metrics are still widely used among researchers and are endorsed by safety regulating organizations such as the National Highway Traffic Safety Administration (NHTSA) \citep{Laituri2016} and the National Operating Committee on Standards for Athletic Equipment (NOCSAE) \citep{Committee2012}. 
More recently, brain tissue deformation-based metrics have been introduced that use head kinematics as input to computational models that can approximate the effect of head motion on brain displacement and deformation. These metrics either use simple discrete mechanical elements in lumped-parameter models, i.e. mass-spring-damper combinations, to give a rigid-body estimate of brain's relative motion with respect to the skull \citep{Kornhauser1954a,Low1987,laksari2015,Gabler2018b}, or more complex finite element (FE) models with detailed geometry of the brain anatomy, which can simulate the local brain deformation and interaction with the stiff bony or membranous structures \citep{Kleiven2013,Zhao2016a,Ji2014c}.  
In the case of lumped models, brain angle metric (BAM), developed based on a data set of concussive and sub-concussive head impacts \citep{Laksari2019}, and in the case of FE models, maximum principal strain (MPS) and axonal fiber strain (FS) along the white matter axon fibers have been proposed as effective injury diagnosis metrics \citep{Wu2019b}. 
\\

Evidently, both for the kinematic-based and the brain deformation-based metrics, head impact kinematics play a major role. With the advent of wearable sensor technology, several groups have been collecting on-field head kinematic measurements during contact sports events \citep{Hernandez2014,Laksari2018,Miller2019a,Wu2018a,Miller2018}. However, despite these pioneering efforts, on-field head kinematic measurements are not widely available. As a result, researchers have resorted to simplifying parameterizations of head collisions as idealized biphasic acceleration impulses \citep{Ji2014b,Abderezaei2019a,Yoganandan2008}. These biphasic impulses are commonly represented either by a triangle or half-sine and defined by two parameters: height and width constitute the magnitude and duration of a head impact impulse. The simplification of kinematic impulses serves the objective of emulating on-field kinematic data of a head impact with a few and manageable number of parameters to populate an otherwise infinite-dimensional loading space to investigate and establish a relation between head motion and brain injury. However, a potential disadvantage of these simplifications is overlooking valuable information that could prove detrimental in developing injury metrics. Therefore, it is paramount to understand the characteristics of real-world head impacts and whether we can accurately capture them through simplified approximations. 
% \\
Furthermore, advances in computational methods, including machine learning algorithms, have provided new and exciting avenues . As a result, given the prohibitively high computational cost of current FE models, the biomechanics community has been trying to utilize such interpolative and machine learning techniques \citep{Ji2014b,Cai2016,Wu2019c,Wu2019b}. However, a limitation of those techniques is the large number of kinematic data required to train these algorithms (in the order of thousands of head impacts \citep{Wu2019c}). Currently such a data set is not widely available. Thus, artificial augmentation of kinematic samples has been utilized as an alternative to satisfy that training data set requirements of such algorithms.
\\

In this paper, we present a formal method to extract the most dominant features of on-field head impact kinematics from an existing data in the context of contact sports. We subsequently use the extracted features in order to construct an augmented data set that resembles the on-field measurements. 
We hypothesize that by using our method, based on principal component analysis (PCA), we will acquire more accurate injury predictions than current biphasic impulse approximations when compared against the ground truth measurements.  
Furthermore, we present a modal reconstruction technique that, despite using relatively few modes, can emulate a desired number of augmented head impacts that are statistically similar to the ground truth impact measurements.

\section*{Materials and methods}

In order to study the characteristics of head impact kinematics and the efficacy of simplified approximations, we used a previously-collected data set of 537 head impact kinematics measured during contact sports, including American football, boxing, and mixed martial arts \citep{Hernandez2014,Laksari2018}. For each impact, 6 degrees of freedom (DoF) kinematics -- linear acceleration and angular velocity in the three anatomical directions -- were collected at 1,000Hz for 100$ms$ using a mouthguard instrumented with a triaxial accelerometer and a triaxial gyroscope \citep{Hernandez2014}. We construct three different reduced kinematics data sets to approximate the measured kinematics: 1) using principal component analysis (PCA), we decrease the dimensionality of the measured head impact kinematics to construct a low-rank kinematics data set, and 2-3) using previously proposed biphasic assumptions for acceleration impulses with acceleration magnitude and duration as the two variables, we construct biphasic data sets once for triangle (Tri) and once for half-sine (HS) approximations. 
We investigate the efficacy of each approximation by comparing its performance in detecting brain motion/deformation and injury prediction using three types of metrics: 1) kinematics-based injury metrics, including HIC, RIC and BrIC, 2) brain angle metric (BAM), and 3) tissue deformation-based finite element injury metrics, including maximum principal strain (MPS) and axonal fiber strain (FS) in the whole brain (WB) and corpus callosum (CC).  
We compare the performance of each approximation against the ground truth (GT) data described above. We perform power spectral density (PSD) analysis on the derived temporal modes to obtain their predominant frequencies. These values are given by the maximum power spectral density (PSD) values of each mode. 
Finally, we present a modal reconstruction method for emulating augmented head impacts.

%===========================================================
%=========================================================== 
\subsection*{Dimension reduction through principal component analysis}

Our goal is to exploit the correlations between different measurements and find a reduced representation for three quantities of interest (QoIs), including: linear acceleration, angular velocity and angular acceleration in each anatomical direction.  In the case of linear kinematics,  anterior-posterior, inferior-superior, and lateral directions are considered and in the case of angular kinematics, axial, coronal and sagittal directions are considered as separate QoIs. To this end, we apply principal component analysis (PCA) to our data set. For each QoI, we form a data matrix $\bm{X}_{m\times n}$, where $n=537$ is the number of measured head impacts and $m=100$ is the number of time steps, and $\bm{X} = [\bm{x}_1 | \bm{x}_2 | \dots | \bm{x}_n ]$, where each column represents the measured QoI for a particular head impact and  each row represents the time instance of the measurement. To perform  PCA, we compute the singular value decomposition (SVD) of the data matrix: $\bm{X}=\bm{U}\bm{\Sigma}\bm{Y}^T$, where $\bm{U}=[\bm{u}_1 | \bm{u}_2 | \dots | \bm{u}_n ]$ are a set of orthonormal modes, i.e. $\bm{u}^T_i \bm{u}_j = \delta_{ij}$, $\bm{\Sigma}=\mbox{diag}(\sigma_1, \sigma_2, \dots, \sigma_n)$  is a diagonal matrix, where $\sigma_1 \geq \sigma_2 \geq \dots \geq \sigma_n$ are the singular values, and $\bm{Y}=[\bm{y}_1 | \bm{y}_2 | \dots | \bm{y}_n ]$ are the uncorrelated linear components, i.e. $\bm{y}^T_i \bm{y}_j = \delta_{ij}$ with  the joint probability distribution function (PDF) of $p(y_1, y_2, \dots, y_n)$. This gives an ordered array of the modal contributions inherent to head impacts response. Finally, we perform power spectral density (PSD) analysis on the derived modes $\bm{u}_i$ to obtain their predominant frequencies. These values are given by the maximum PSD values of each mode. 

\vspace{3mm} 
A reduced representation of the impact data is obtained by:
$\bm{X} \simeq \sum_{i=1}^k \sigma_i  \bm{u}_i \bm{y}^T_i $.  To quantify the performance of the reduction,  we introduce: 

\begin{equation}
\eta(k) =\frac{(\sum_{i=1}^k \sigma_i)}{( \sum_{i=1}^n \sigma_i)}, 
\label{eq:svcriteria}
\end{equation}

\subsection*{Data-Driven emulator}

Once we extract the modal characteristics of the head impact data set, a reduced emulator of the head impact kinematics is obtained by truncating to $k$ PCA modes:
\begin{equation}\label{Eq:emulator}
\bm{x}^* = \sum_{i=1}^k \sigma_i y_i^* \bm{u}_i,
\end{equation}
where $(y^*_1,y^*_2, \dots, y^*_k)$ is a random point with $k$ components drawn from the marginal PDF of $p(y_1, y_2, \dots, y_k)$. Equation \ref{Eq:emulator} can be used as an emulator for producing new time series ($\bm{x}^*$) for each of the QoIs that are nearly indistinguishable from the ground truth head impact kinematics measurements. Our approach can be interpreted as a stochastic dimension reduction technique and it is a special case of dimension reduction with time-dependent modes \citep{SL09,Babaee_PRSA,Babaee:2017aa,B19,PB20}, in which the loading is a random input and the QoIs are random output.  In order to ensure high quality augmented data, we establish features to statistically compare new time series with the GT, including parameters such as impact time, peak and duration of impacts, and orthonormal projections of PCA modes. 
We will provide this emulator as a stand-alone program that allow users to build low-rank head kinematics data sets with various approximation levels ($k$) and emulate a pre-defined number of impacts.

\begin{figure}%[b]
    \centering
    \includegraphics[width=0.9\textwidth]{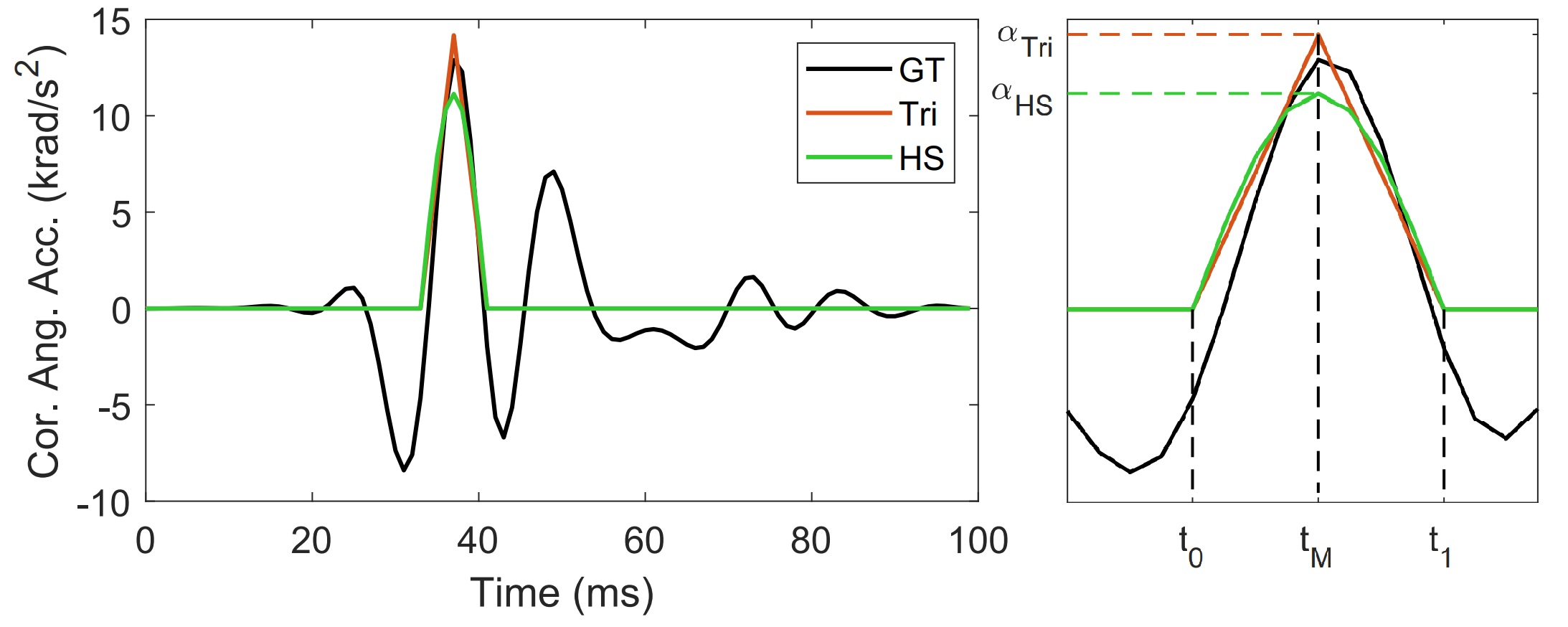}
    \caption{Constructing triangular and half-sine biphasic impulses \citep{Ji2014b,Abderezaei2019a}: a representative angular acceleration trace in the coronal plane, magnified to better represent initiation time ($t_0$), completion time ($t_1$), time of peak ($t_M$), and peak acceleration ($\alpha_{Tri}$ and $\alpha_{HS}$).} 
    \label{fig:buildtrisin}
\end{figure}

\subsection*{Parametrizing biphasic impulse profiles}

In order to compare the performance of previously proposed biphasic models for angular and linear acceleration impulses \citep{Ji2014b,Abderezaei2019a} against our low-rank PCA approximation, we reconstruct biphasic triangle (tri) and half-sine (hs) representations of the 537 head impacts described above. First, the maximum absolute value of the angular acceleration profile ($\alpha_M$) -- computed by differentiating angular velocity measurements using a first-order forward divided difference method (two points) -- was identified, including the time of peak ($t_M$). The impact duration ($\Delta t$) is defined as the time interval on either side of $t_M$. The boundary of this interval is defined as where the sign or the convexity of the acceleration profile (whichever comes first) changes. Convexity changes are computed based on the second derivative test using a common three-point stencil central finite difference derivative. This process is described in Figure \ref{fig:buildtrisin}, where the impact duration ($\Delta t=t_1-t_0$) is the time elapsed between the initiation time ($t_0<t_M$) and the completion time ($t_1>t_M$) of impact. In the cases where $t_M\rightarrow 0ms$ or $t_M \rightarrow 100ms$, since it is not possible to define $t_0$ or $t_1$ correctly, only half of the simplified pulse was created. Finally, the change in velocity ($\Delta\omega$) was computed as the area under the acceleration impulse ($\Delta\omega=\int_{t_0}^{t_1} \alpha(t) dt$), and the corresponding acceleration magnitudes for the triangle ($\alpha_{tri}$) and half-sine ($\alpha_{hs}$) approximations were calculated through:

        \begin{equation}
             \displaystyle %\begin{cases}
             \alpha_{tri}=2\cfrac{\Delta\omega}{\Delta t},
            %  \\
            \hspace{15mm}
             \alpha_{hs}=\cfrac{\pi\Delta\omega}{2\Delta t}.
            %  \end{cases}
        \label{netangular}
        \end{equation}
        
Angular velocity pulses were then computed through direct temporal integration of low-rank angular acceleration pulses.

%===========================================================
%===========================================================    
\subsection*{Accuracy of injury prediction metrics}

\subsubsection*{Kinematics-based injury metrics:}

We used HIC$_{15}$, RIC$_{36}$ and BrIC to compare the performance of our proposed PCA reduction against the triangle and half-sine  biphasic signals, i.e. triangle and half-sine approximations. To this end, we used previously published injury threshold values: 1) for HIC$_{15}$, values of $240$ and $667$ have been reported as $50\%$ risk of concussion \citep{Newman2000c} and skull fracture \citep{Marjoux2008b}, respectively; 2) for RIC$_{36}$, a value of $10.3\times 10^6$ is reported as $50\%$ risk of concussion \citep{Kimpara2012a}; and 3) for BrIC, a value of 0.5 constitutes a $50\%$ concussion risk \citep{Takhounts2013}. We used these thresholds to assess the performance of each reduction approach in providing injury predictions in terms of sensitivity and specificity with respect to the ground truth measurements. 

Injury thresholds were used as indicators, defining true positives (above the threshold) and negatives (below the threshold), while the predictive value of each impulse approximation was compared against the GT measurement.

\begin{figure}[t]
    \centering
    \includegraphics[width=\textwidth]{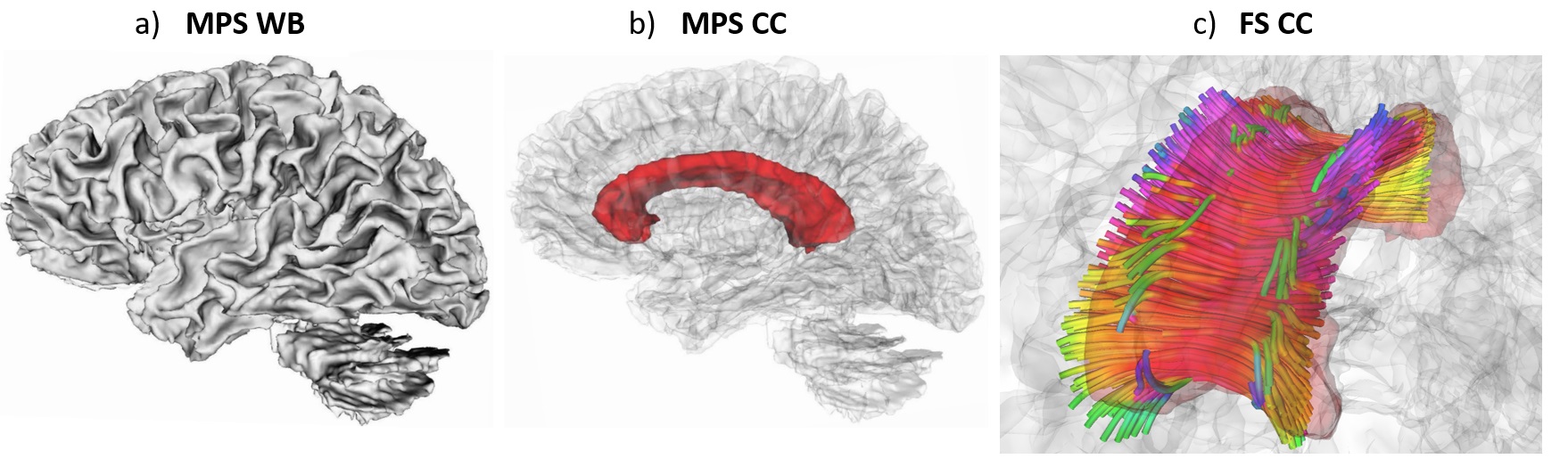}
    \caption{Representation of strains metrics with highlighted regions: a) maximum principal strain (MPS) in the whole brain (WB), b) maximum principal strain (MPS) in the corpus callosum (CC), c) fiber strain (FS) in the corpus callosum (CC). Fiber colors represents directions: inferior--superior (blue), lateral (red) and anterior--posterior (green). Images were generated using 3DSlicer from ATLAS-based anatomical representation in FreeSurfer \citep{Zhang2018}.}
    \label{brains}
\end{figure}

%===========================================================
%===========================================================
\subsubsection*{Brain angle injury metric:}

We further compared the performance of each approximation using injury criterion based on  lumped-parameter models of the head. 
These models generally consider simplifying assumptions: skull and brain are considered rigid bodies and relative motion between the two represents a form of deformation and injury, and the compliance of the brain-skull interface such as the effect of bridging veins, dura and pia maters is represented by linear spring and damper elements \citep{Kornhauser1954,Low1987,Gabler2018}. Given the head kinematics as the base excitation input, these models can estimate the relative motion of brain and skull, particularly the angular motion since that has been seen as the more consequential type of motion \citep{Sullivan2015}. Recently, brain angle metric (BAM) was developed based on the characteristics of human brain and skull in finite element simulations, and validated against observed concussive and sub-concussive head impacts \citep{Laksari2019}. We compute BAM for each kinematic approximation (PCA, triangle and half-sine).

%===========================================================
%===========================================================
\subsubsection*{Tissue deformation-based injury metrics:}
As a final step in studying the efficacy of the different kinematic approximations, we compared the performance of each approximation in predicting the tissue-level deformation metrics using finite element simulations, including maximum principal strain (MPS) in the whole brain (WB) and in the corpus callosum (CC) region, as well as axonal fiber strains (FS) in the corpus callosum region, which have all been proposed as predictive tissue-level metrics for injury classification \citep{Kleiven2007a,Ji2014c,Laksari2018} (Figure \ref{brains}). Recently a convolutional neural network (CNN) was developed based on pre-trained FE simulations based on the Worcester Head Injury Model (WHIM) \citep{Zhao2017}. This CNN method uses angular velocity data as input to approximate the regional brain deformations, i.e. maximum principal and axonal fiber strain \citep{Wu2019c}.

%===========================================================
%===========================================================
\subsubsection*{Injury metric error analysis:}
Having simulated the injury metrics for each head impact measurement, we calculated the corresponding injury metric (metric$_{GT}$) and the injury metric estimated by the kinematics approximation (metric$_{approx}$) using the equation below:
    \begin{equation}
        error=\left|\frac{\text{metric}_{approx}-\text{metric}_{GT}}{\text{metric}_{GT}}\right|\times100.
        \label{BAMerror}
    \end{equation}
    
\vspace{0.1cm} 
\noindent    
Subsequently, we performed Friedman test \citep{daniel1990applied} with a $p$ value of 0.01 (MATLAB, \verb|friedman|) to show significant differences between each approximated impulse and the ground truth. We also performed sensitivity and specificity analysis to provide an estimate for the efficacy of approximating the metrics for injury diagnosis. For this purpose, we used previously published values for 50\% risk of concussion, including MPS$_{WB}=0.2$ \citep{Patton2012a}, MPS$_{CC}=0.2$ \citep{Kleiven2007a}, and FS$_{CC}=0.074$ \citep{Giordano2014b}.

\begin{figure}[h]
    \centering

    \includegraphics[scale=0.47]{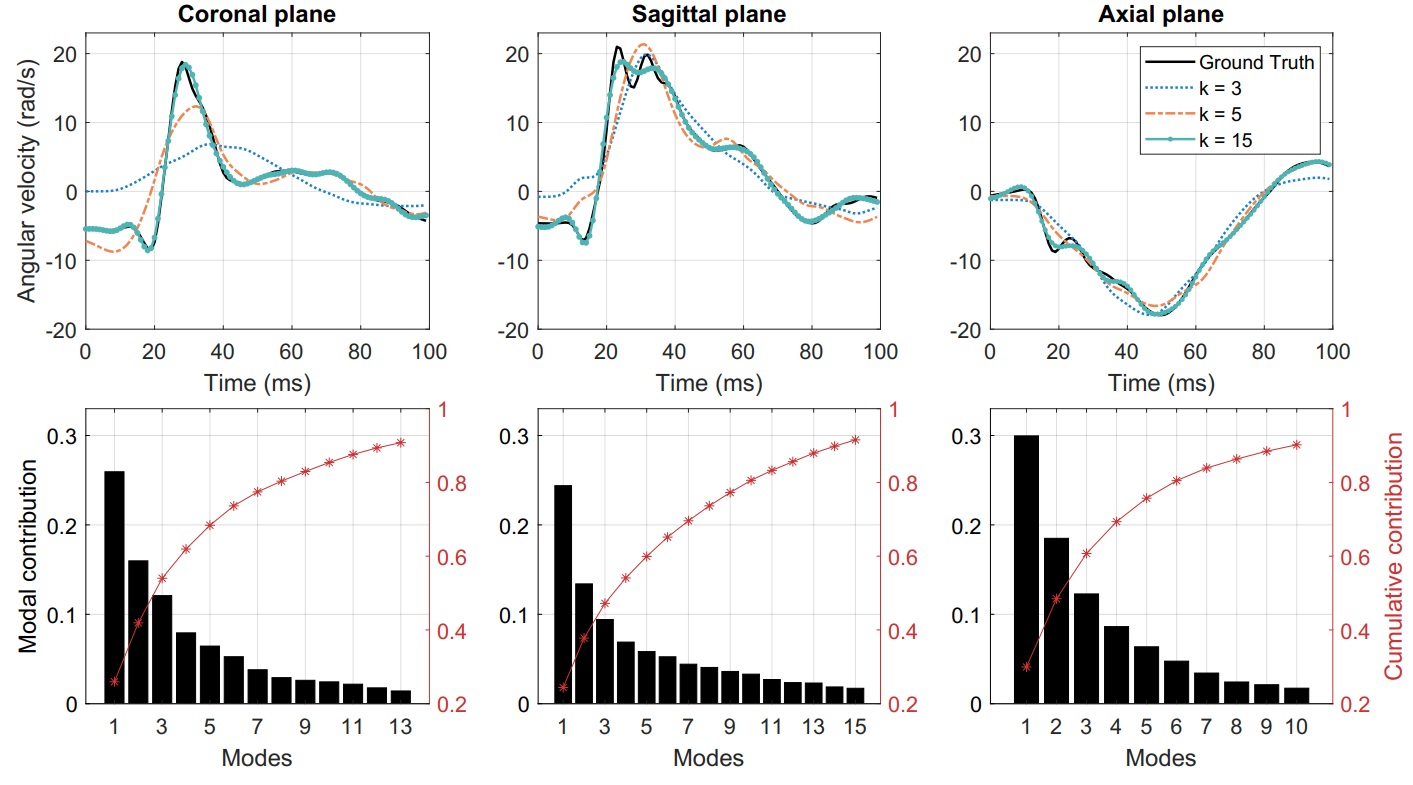}
    \caption{{\it Top:} Low-rank reconstruction of angular velocity using $k=3, 5, 15$ PCA modes. {\it Bottom:} Individual and cumulative contribution of PCA modes for angular velocity reconstruction. Columns from left to right show results for coronal, sagittal and axial directions, respectively.}
    \label{fig:pod} 
\end{figure} 

%===========================================================
%===========================================================
\section*{Results}

\subsection*{Dimension reduction through PCA}

We performed PCA on the measured kinematics data for the QoIs, i.e. linear acceleration, angular velocity and angular acceleration in each anatomical direction.  We used the reduction criterion of $\eta = 0.90$, as defined in Equation~(\ref{eq:svcriteria}), for all these cases. In the case of angular velocity, the minimum number of modes that satisfies this reduction criterion is $k=13$, 15 and 10 modes for coronal, sagittal and axial directions, respectively. 
In Figure~\ref{fig:pod} (top row),  the PCA reconstruction of angular velocity in three anatomical directions for a sample case is shown. The sample case was chosen randomly from the 537 cases and it is represented by a column of the data matrix $\bm{X}$. The ground truth measurement for the sample case as well as the reconstructed impulses with different levels of reduction are shown. It is clear that the 15-mode reduction yields a satisfactory reconstruction.  In Figure~\ref{fig:pod} (bottom row) the individual and cumulative contribution of PCA modes are shown for the entire kinematics data set. These results demonstrate that with a relatively small number of PCA modes an accurate approximation of the head kinematic measurements can be achieved. As an additional analysis, in order to determine the convergence of this method, we performed PCA with several randomly selected subsets of the 537 ground truth measurements (with 100, 200, 300, 400, and 500 samples) and investigated the number of modes required to satisfy the $\eta>0.90$ criterion. The results show that the minimum number of modes slightly grows with subsets size but levels off below 500 cases, indicating that our data set of 537 could be sufficient to reliably reconstruct a head impact data set (See SM.8).

\subsection*{Data-driven emulator}

\begin{figure}[h]
    \centering
    \includegraphics[width=0.9\textwidth]{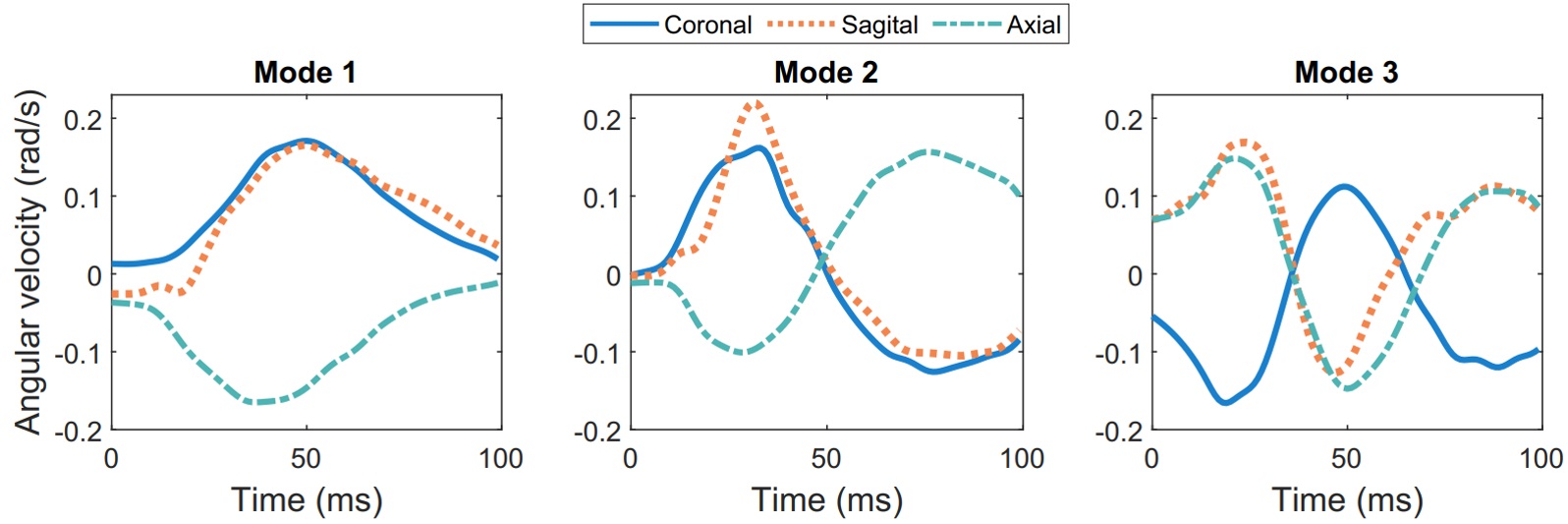}
    \caption{The three most energetic temporal modes for angular velocity for the entire 537 head impact data set.}
    \label{fig:tmodes}
    
\end{figure}

\begin{figure}[h]
    \centering
  \includegraphics[width=\textwidth]{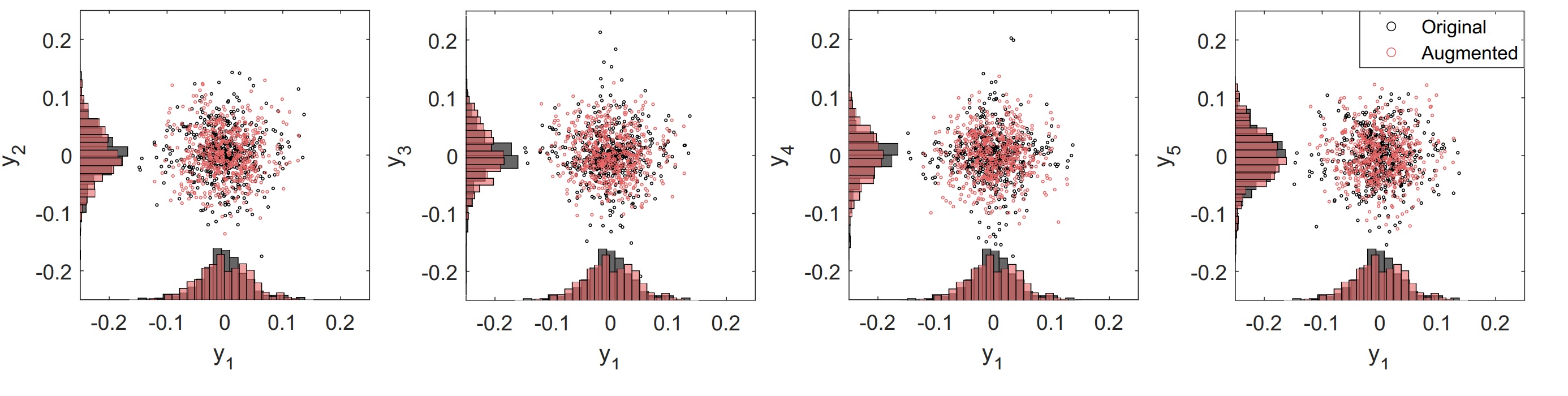}
\caption{Distribution graphs for second to fifth principal components ($\bm{y_{2},...,\bm{y_{5}}})$ projected on to the first principal component ($\bm{y_1})$ for angular velocity in the sagittal direction. PCA data (in black) follows a Gaussian distribution. Emulated data (red) is generated performing a Gaussian random number generator based on mean and variances from the PCA modes. }
\label{scatter}
\end{figure}

The first three temporal modes show a classic modal behavior with 1, 2, and 3 peaks in all three anatomical directions (Figure \ref{fig:tmodes}). 
Together, these first three modes capture nearly half of the total angular velocity response, and with each additional mode, we can reconstruct a closer approximation with the ground truth. For more analysis of the modes, see Supplementary Materials Figures SM.2 and SM.4.
To further study the distribution of these modal approximations, we 
show the orthonormal projection of the first five PCA modes ($y_2,...,y_5$) against the first and most energetic mode ($y_1$) of the PCA data (black circles and bars in Figure \ref{scatter}). It is clear that the modes follow a Gaussian distribution, which would be an important consideration for emulating more data points.  

To show the performance of our data-driven emulator, we reproduced an additional $k=$537 head impact cases by randomizing the $y_k^*$ columns each with the same mean and variance as the original $\bm{Y}$ matrix (Equation \ref{Eq:emulator}). Projection of the orthonormal vectors $\bm{y}^*_i$ on $\bm{y}^*_1$ for an augmented data set (red circles and bars in Figure \ref{scatter}) shows similar distribution as the ground truth data. In addition, features of the augmented data generated on our emulator, such as duration and peak distributions, have not significant differences with respect those of GT (See figures SM.5, SM.6 and SM.7). 
Thus, our emulator is able to successfully generate reliable kinematics sets of data based on real on-field measurements. A copy of the head impact kinematics emulator will be available on our website \citep{EmulatorLink}.

\subsection*{Natural frequencies}

\begin{figure}[h]
    \centering
    \includegraphics[width=\textwidth]{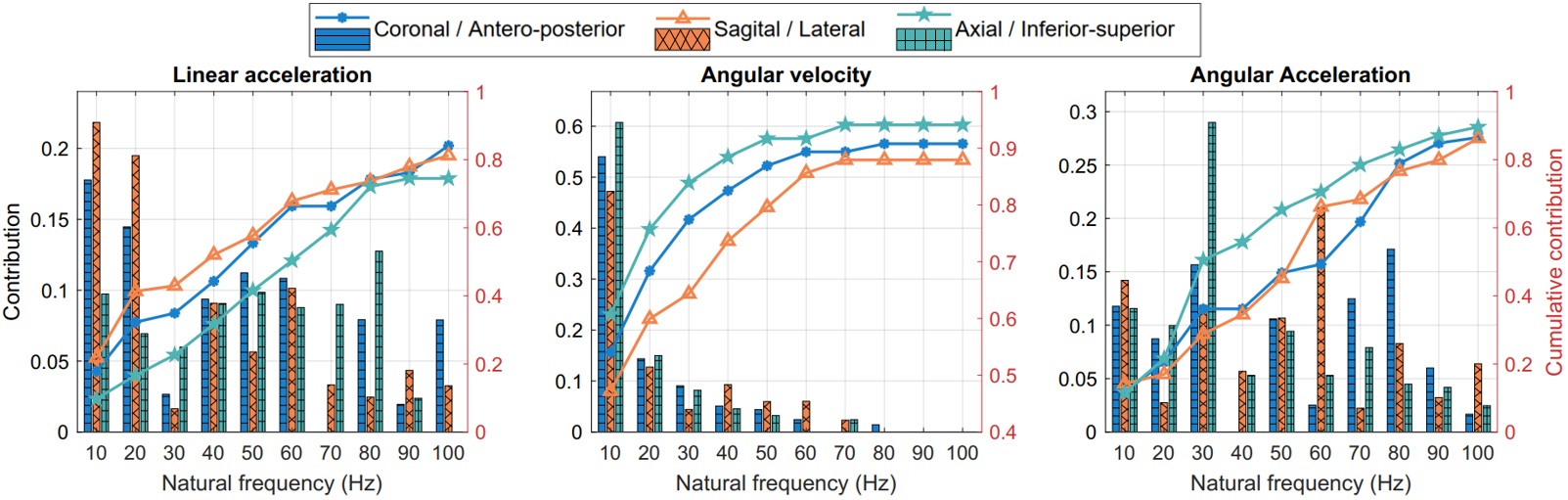}
    \caption{Natural frequencies and their contribution to head motion kinematics.} 
    \label{fig:nfreq}
\end{figure}

The contribution of the most dominant frequencies, obtained through PSD criterion, are displayed in Figure \ref{fig:nfreq}. In general, low frequencies interval such from $10$ to $40Hz$ have the highest predominance for each parameter. Notably, the cumulative contribution for rotational velocity is progressively decreasing with increasing frequency.

\subsection*{Parametrizing biphasic impulse profiles}
Using the criteria described above, we fitted triangle and half-sine analog pulses to the ground truth kinematics measurements in order to parametrize the rotational acceleration magnitude and duration for each head impact. As a result, we derived 537 analog impulses in the three anatomical directions for both triangle and half-sine approximations. The results are presented in Table \ref{table:trisinmetrics}, where the rotational acceleration magnitude and duration average and standard errors of the mean are given for the ground truth and the impulses approximations.

\begin{table}[ht]
    \centering
    
    \caption{Mean and standard error of acceleration magnitude and duration for ground truth and three approximations (PCA, triangle, half-sine). The PCA magnitudes were obtained by decomposing the ground truth angular accelerations for the criterion $\eta=0.90$, which constituted of 21 modes for coronal and sagittal directions and 20 modes for axial direction (See Figure SM.3).} 
    \begin{adjustbox}{width=1\textwidth}
    
    \begin{tabular}{lcccc}
    \hline
          & \textbf{Coronal direction} & \textbf{Sagittal direction} & \textbf{Axial direction} \\ \hline
    \textbf{Ground truth magnitude $(rad/s^2)$ } &  818.89 $\pm$ 937.12 & 1,498.10 $\pm$ 1,753.40 & 655.16 $\pm$ 535.03 \\ %\hline
    \textbf{PCA magnitude $(rad/s^2)$ } &  801.41 $\pm$ 922.08 & 1,460.50 $\pm$ 1,736.10 & 641.52 $\pm$ 523.06 \\ %\hline
    \textbf{Triangle magnitude $(rad/s^2)$} & \hspace{0.0cm} 871.53 $\pm$ 1,011.8 & 1,625.40 $\pm$ 1,923.20 & 697.04 $\pm$ 553.95 \\ %\hline
    \textbf{Half-sine magnitude $(rad/s^2)$}  & 681.25 $\pm$ 790.38 & 1,269.30 $\pm$ 1,500.40 & 545.64 $\pm$ 433.40 \\ %\hline
    \textbf{Duration $(ms)$} & \hspace{-0.35cm} 15.20 $\pm$ 6.75 & \hspace{-0.3cm} 15.00 $\pm$ 8.03 & \hspace{-0.3cm} 17.90 $\pm$ 8.45 \\ \hline
    \end{tabular}
    \end{adjustbox}
    
    \label{table:trisinmetrics}
\end{table}

\subsection*{Accuracy for injury prediction metrics}

\subsubsection*{Kinematic-based injury metrics}

\begin{figure}[h]
    \centering
    \includegraphics[width=\textwidth]{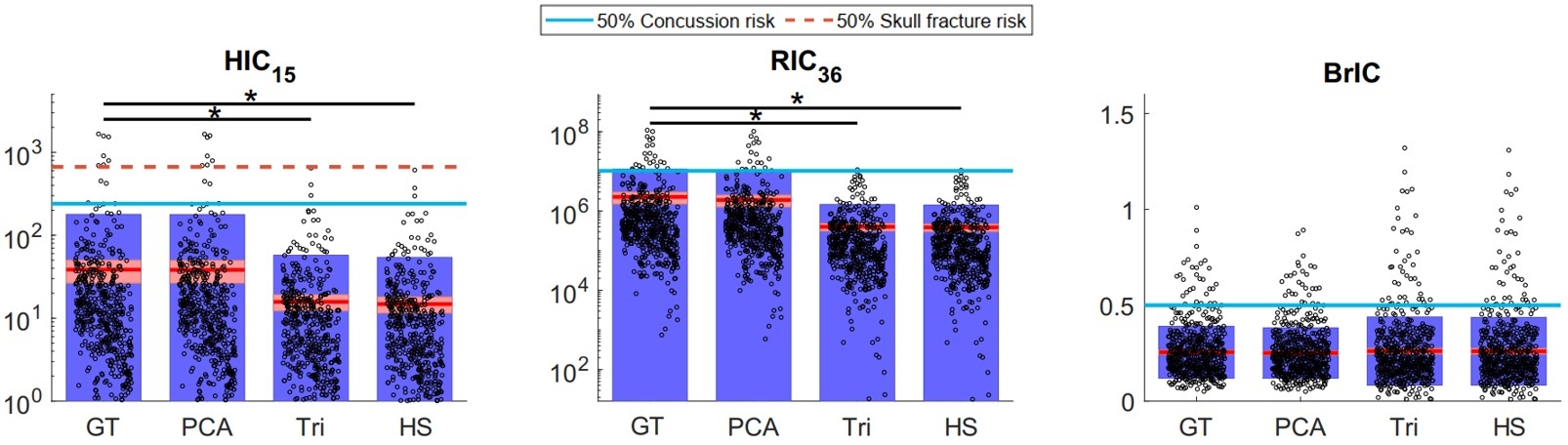}
    \caption{Computed HIC$_{15}$, RIC$_{36}$ and BrIC for each of the ground truth (GT) and the three approximated data sets: PCA, triangle (Tri), and half-sine (HS). The circles represent each sample, the solid red line represents the mean, and the blue and red regions show the standard deviation and standard error, respectively. The solid blue and dashed red lines represent $50\%$ risk of concussion and skull fracture, respectively. Significant differences are indicated for $p<0.01$. HIC and RIC graphs are on log-scale.}
    \label{hicricbric}
\end{figure}

The injury metrics HIC$_{15}$, RIC$_{36}$ and BrIC were computed for every model and the ground truth. Figure \ref{hicricbric} shows the distribution of all samples and the corresponding concussion and skull fracture thresholds. 
The PCA predictions showed similar mean and standard deviations for HIC and RIC ($38.20\pm139.55$ and $1.89\times 10^6\pm7.66\times 10^6$ respectively) as the ground truth ($38.49\pm140.13$ and $2.26\times 10^6\pm9.04\times 10^6$); however, there is a significant difference between the ground truth predictions and the triangle ($15.74\pm41.71$ and $3.98\times 10^5\pm1.06\times 10^7$) and half-sine ($14.80\pm39.31$ and $3.86\times 10^5\pm1.02\times 10^6$) impulses. 

Additionally, whereas the PCA-based impulses showed accurate predictions compared to the ground truth, we observed that the biphasic approximations either under-predicted injury (higher number of false negatives) in terms of HIC and RIC, or over-predicted injury (higher number of false positives) in terms of BrIC (Figure \ref{hicricbric}). 

To better illustrate this, we performed sensitivity and specificity analysis for injury classification with respect to the ground truth, where the PCA-based signals showed high predictive performance compared to the biphasic impulses (Table \ref{hicricbricsensspec}).

\begin{table}[h]
\caption{Sensitivity and specificity of kinematics-based injury metrics for HIC$_{15}$, RIC$_{36}$ and BrIC for the three approximations: PCA-based method, triangles (Tri) and half-sine (HS) compared to the ground truth, considering thresholds of 50\% risk of injuries.}
\begin{adjustbox}{width=1\textwidth}
\begin{tabular}{  l c p{1cm}p{1cm}p{1cm} p{0.5cm} p{1cm}p{1cm}p{1cm} p{0.5cm} p{1cm}p{1cm}p{1cm} p{0.5cm} p{1cm}p{1cm}p{1cm}  }
\hline
 & & \multicolumn{3}{c}{HIC} & & \multicolumn{3}{c}{HIC} & & \multicolumn{3}{c}{RIC} & & \multicolumn{3}{c}{BrIC} \\
 & & \multicolumn{3}{c}{(50\% risk -- skull fracture)} & & \multicolumn{3}{c}{(50\% risk -- concussion)} & & \multicolumn{3}{c}{(50\% risk -- concussion)} & & \multicolumn{3}{c}{(50\% risk -- concussion)} \\
 \cline{3-5} \cline{7-9} \cline{11-13} \cline{15-17}
 & & PCA & Tri & HS & & PCA & Tri & HS & & PCA & Tri & HS & & PCA & Tri & HS \\
\hline
Sensitivity & & 1.00 &  0.00 &  0.00 & & 0.91 & 0.27 & 0.27 & & 0.79 & 0.04 & 0.04 & & 0.9 & 0.77 & 0.77 \\ %\hline
Specificity & & 1.00 & 1.00 & 1.00 & & 1.00 & 1.00 & 1.00 & & 1.00 & 1.00 & 1.00 & & 1.00 & 0.97 & 0.97 \\
\hline
\end{tabular}
\end{adjustbox}
\label{hicricbricsensspec}
\end{table}

\subsubsection*{Brain angle injury metric:}

\begin{figure}[h]
    \centering
    \includegraphics[width=\textwidth]{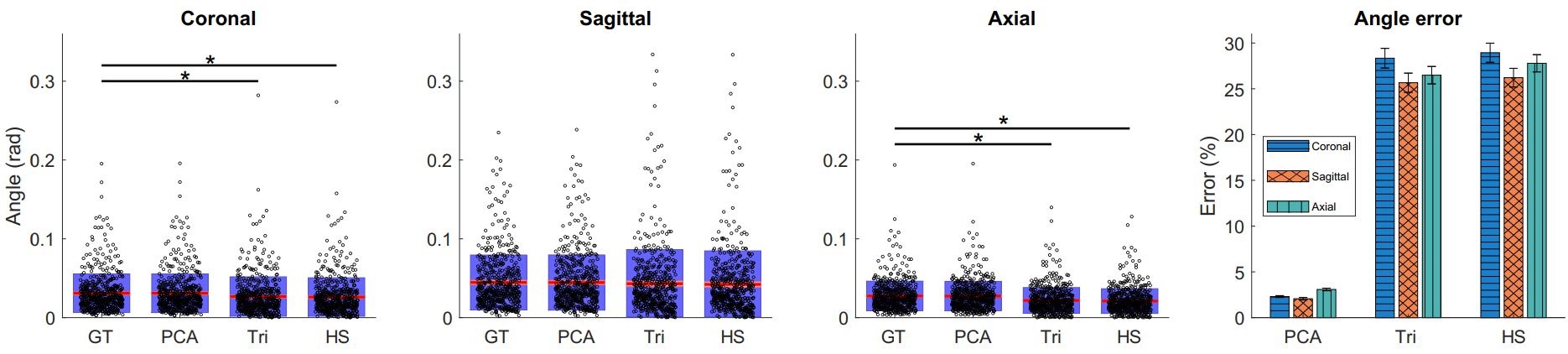}
    \caption{Computed brain angle metric (BAM) values for each impact case based on ground truth and the corresponding PCA and biphasic approximations. The significant differences are indicated for $p<0.01$. The error comparison between models with means and SEM of each data set is shown on the right.}
    \label{fig:angleerror}
\end{figure}

We computed 3DoF relative brain angles using the lumped model proposed in \citep{Laksari2019} to obtain the maximum resultant relative brain angle as a result of each head impact based on the ground truth kinematics and each of the three approximations. In coronal and axial directions, triangular and half-sine approximations gave significantly lower predictions for the brain angle metric, whereas the PCA modes showed no statistically significant difference from the ground truth (Figure \ref{fig:angleerror}). 
We observed substantially smaller approximation errors (Equation \ref{BAMerror}) for the PCA approach ($\sim3\%$) compared to the two biphasic approximations ($\sim25\%$) (Figure \ref{fig:angleerror}). 

\subsubsection*{Tissue deformation-based injury metrics:}

\begin{figure}[h]
    \centering
    \includegraphics[width=\textwidth]{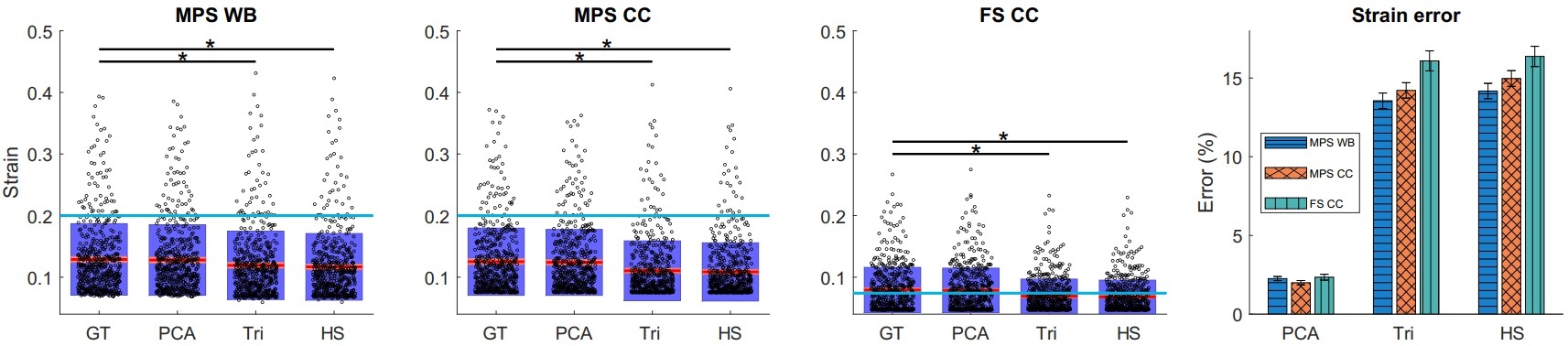}
    \caption{Simulated strain metrics for ground truth head impact measurements and the corresponding PCA and biphasic approximations. The blue line shows the $50\%$ risk of concussion for each metric. The significant differences are indicated for $p<0.01$. Also the mean and standard error of the mean are shown on the right for strain estimation errors. }
    \label{strains}
\end{figure}

Similar to brain angle metric, we calculated the errors between the ground truth strain metrics and those of low-rank approximations. 
As can be seen in Figure \ref{strains}, the PCA approach closely follows the ground truth simulation results in all three strain metrics: maximum principal strain (MPS) in the whole brain (WB) and corpus callosum (CC), as well as axonal fiber strain (FS) in the corpus callosum. The strains computed for the biphasic impulses significantly differ from the ground truth values and successively provide higher errors as we include region-specific and morphological information. Furthermore, based on the 50\% concussion risk thresholds mentioned above, the PCA approach provides substantially higher sensitivity and specificity for injury classification (Table \ref{strainsensspec}). 

\begin{table}[h]
\centering
\caption{Sensitivity and specificity of strain metrics for maximum principal strain (MPS) for whole brain (WB) and corpus callosum (CC), as well as fiber strain (FS) for CC compared to the ground truth, with respect the threshold of 50\% risk of concussion.}
\begin{adjustbox}{width=.85\textwidth}
\begin{tabular}{  l c p{1cm}p{1cm}p{1cm} p{0.5cm} p{1cm}p{1cm}p{1cm} p{0.5cm} p{1cm}p{1cm}p{1cm}  }
\hline
 & & \multicolumn{3}{c}{MPS WB} & & \multicolumn{3}{c}{MPS CC} & & \multicolumn{3}{c}{FS CC} \\ \cline{3-5} \cline{7-9} \cline{11-13}
 & & PCA & Tri. & H.S. & & PCA & Tri. & H.S. & & PCA & Tri. & H.S. \\
\hline
Sensitivity & & 0.93 & 0.74 & 0.63 &  & 0.98 & 0.55 & 0.53 & & 0.97 & 0.63 & 0.62 \\
Specificity	& & 0.99 & 0.99 & 0.99 &  & 0.99 & 0.99 & 0.99 & & 0.99 & 0.94 & 0.95 \\
\hline
\end{tabular}
\end{adjustbox}
\label{strainsensspec}
\end{table}

\section*{Discussion}

In this study we provide a formal approach for reducing the dimensionality of head impact kinematics in contact sports settings. We first derived the most important modes contributing to the head kinematics through principal component analysis and then compared those to existing methods that approximate head kinematics with simple biphasic impulses. We show that the modal decomposition approach can capture the kinematic behavior of the head with better accuracy and provide better approximations of brain deformation and injury classification. This analysis confirms that although head kinematics during head collisions span a wide range of magnitudes and frequencies \citep{Laksari2018}, we can accurately capture the impact head kinematics by using only a relatively small number of modes. 
The low-rank database constructed based on PCA analysis require only 15 modes to build the ground truth angular velocity kinematics with over 90\% accuracy as well as accurately capture the predictive value of head impact kinematics using a variety of injury metrics. 

A major advantage of our approach is that with the acquired modes above, we are able to emulate a head impact data set with any given number of cases without needing access to the ground truth measurements. This emulated data set would closely replicate head impacts measured by on-field wearable sensors that constitute current state of the art. 

\vspace{3mm}
The advantage of this low-rank emulator, in addition to its computational efficiency, is that it avoids simplifying assumptions for the shape of acceleration impulses and only uses empirical measurements. In contrast, the conventional biphasic assumption for head impacts as simple impulses with only two variables, i.e. acceleration magnitude and duration, falls short in providing accurate estimates. This effect is more pronounced for acceleration impulses that are more variable due to the time derivative but is true for velocity profiles as well. 
This apparent lack of accuracy in injury prediction in the biphasic approximations might be due to the fact that the biphasic triangle and half-sine signals are built using acceleration signals and then integrated to give the corresponding velocity profiles. Since there is no restitutive deceleration for these impulses, angular velocity eventually becomes constant after the  acceleration returns to zero, contrary to the actual measured impulses \citep{Yoganandan2008}. In the case of kinematics-based injury metrics, the discrepancies in misidentifying concussive and subconcussive cases by HIC and RIC could be explained by these metrics' dependence on the shape of the acceleration impulse. In contrast, BrIC is only a function of peak angular velocity, and therefore exhibits less sensitivity to the shape of the impulse and  the biphasic impulse's lack of restitutive deceleration (Figure \ref{hicricbric}).
\\

In the case of brain angle metric, the biphasic impulse approximations show over five-fold errors compared to  PCA-based impulses. This difference could be attributed to the simplification of the biphasic models that influences the solution of the mechanical lumped-parameter models. This discrepancy seems to affect the coronal direction the most and the sagittal direction the least for the biphasic approximations. 

Similar to the brain angle results, brain finite element strains showed superior performance by our PCA-based approach. In previous publications, it has been shown that the closer the model is to the correct anatomical and morphological attributes of the brain, the more accurate the injury predictions become \citep{Giordano2014,Zhao2019}. Similarly, we see a decline in performance for the biphasic models as we include more region-specific and morphological details: 
from maximum principal strains (MPS) in the whole brain to MPS in corpus callosum and on to axonal fiber strains (Table \ref{strainsensspec}). 

\vspace{3mm}
Several other points are worth noting based on our analysis. Relative with time domain, we observe differences in kinematics profiles in the three anatomical directions. It seems that the head experiences higher linear accelerations in the anterior-posterior direction (3.54$\pm$2.97 $m/s^2$) compared to lateral (2.62$\pm$2.25 $m/s^2$) and inferior-superior (3.01$ \pm$3.44 $m/s^2$) directions, and higher angular accelerations in the sagittal direction (190.53$\pm$177.34 $rad/s^2$) compared to coronal (314.50$\pm$401.53 $rad/s^2$) and axial (177.34$\pm$128.16 $rad/s^2$) directions. This observation might be attributed to the type and direction of loading in the specific activity, e.g. direction of tackling in football, as well as anatomical features such as the neck constraint in those directions \citep{Eckersley2017}. In the frequency domain, there is a dominant low-frequency response ($10$ to $40 Hz$) in the head kinematics, expressed by $\sim90\%$ of the total angular velocity response in the axial plane, $\sim83\%$ in the coronal plane and $\sim74\%$ in the sagittal plane, confirming previous findings on frequency dependence of head impacts \citep{Wu2015,Laksari2018}. These results could prove useful for designing better helmets and other safety devices to avoid brain injury by targeting specific low-frequency range of motion. 
\\

In summary, our proposed PCA decomposition approach not only provides a deeper understanding of the head's response during impacts, but also provides a formal basis for reconstructing and augmenting head impact kinematics data. 
Our current emulator is built upon the 537 measured on-field head impacts described above and successfully generates new kinematic data, whose features such peak or duration are nearly indistinguishable (see figure SM.7). It is expected that with more on-field measurements, we would be able to improve the performance of the emulator even further, but our convergence analysis showed the available 537 cases to be sufficient (Figure SM.8). This type of approach might prove necessary given the increased need for larger training data sets in modern machine learning algorithms.

%===============================================================
%===============================================================
\section*{Acknowledgment}
We thank Prof. David Camarillo of Stanford University for providing the head kinematics. PA thanks the Fulbright and CONICYT project Equal Opportunities Scholarship 56170002, and also Carissa Grijalva of the University of Arizona, for helping with the visualizations. KL thanks BIO5 Institute at the University of Arizona for partial support for this study under award number 1119000-KL2. We also thank the National Institutes of Health for partial support of this study under grant number R03NS108167

% ##############################################
% ##############################################

\newpage

\bibliographystyle{apalike}

\bibliography{lib.bib} %,Hessam.bib
%\bibliography{OutlinePOD.bbl}
%\input{OutlinePOD.bbl}
%\bibliographystyle{plain}

% ##############################################
% ##############################################

\newpage
\section*{Supplementary Materials}

\setcounter{figure}{0} \renewcommand{\thefigure}{SM.\arabic{figure}} 

\setcounter{table}{0} \renewcommand{\thetable}{SM.\arabic{table}} 

\setcounter{equation}{0} \renewcommand{\theequation}{SM.\arabic{equation}}

\begin{figure}[h]
    \centering

    \includegraphics[scale=0.47]{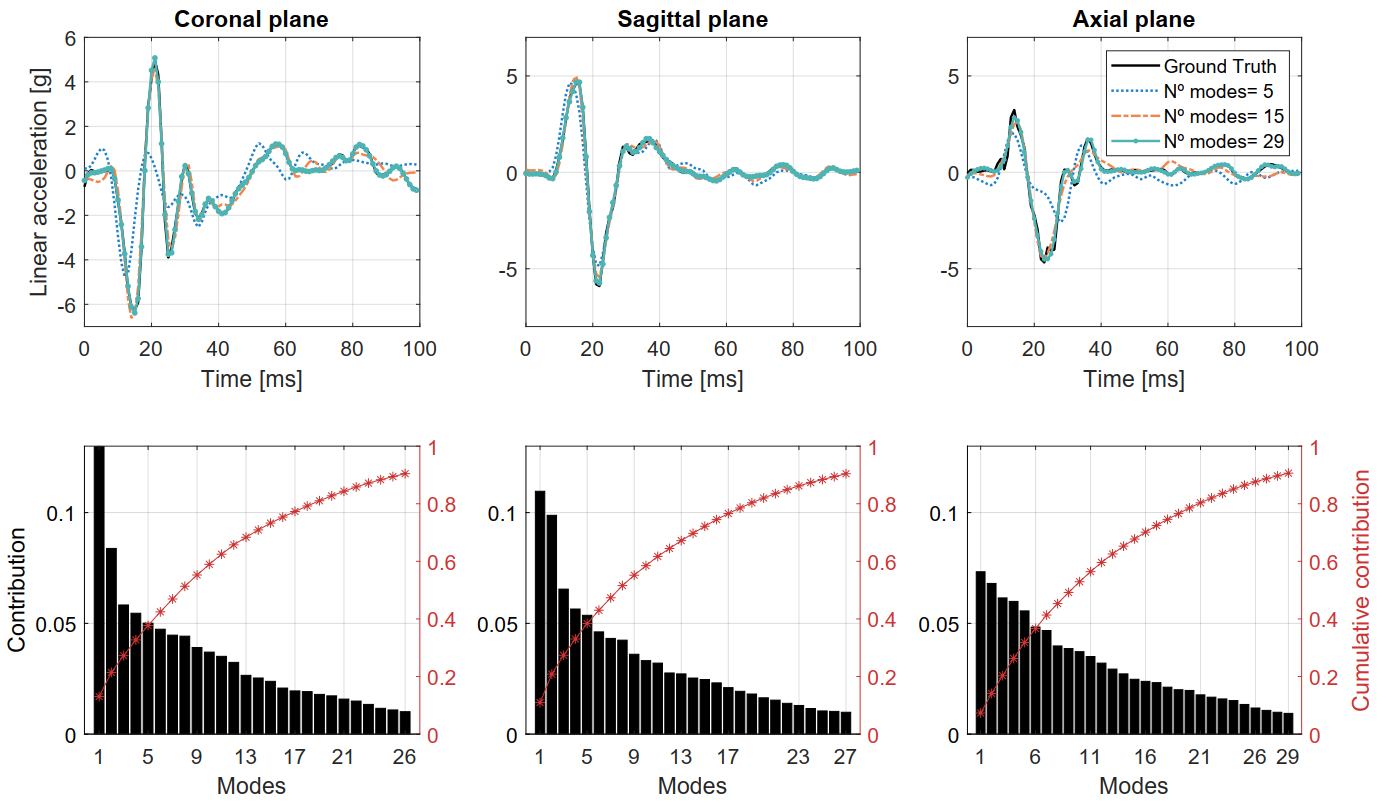}
    \caption{{\it Top:} Representative reconstruction of the linear acceleration data for a single head impact. {\it Bottom:} Individual and cumulative contribution of the temporal modes derived through PCA for angular velocity in each anatomical direction, constituting over 90\% of the ground truth measurements.}
    \label{fig:poda} 
\end{figure}

\begin{figure}[h]
    \centering
    \includegraphics[scale=0.4]{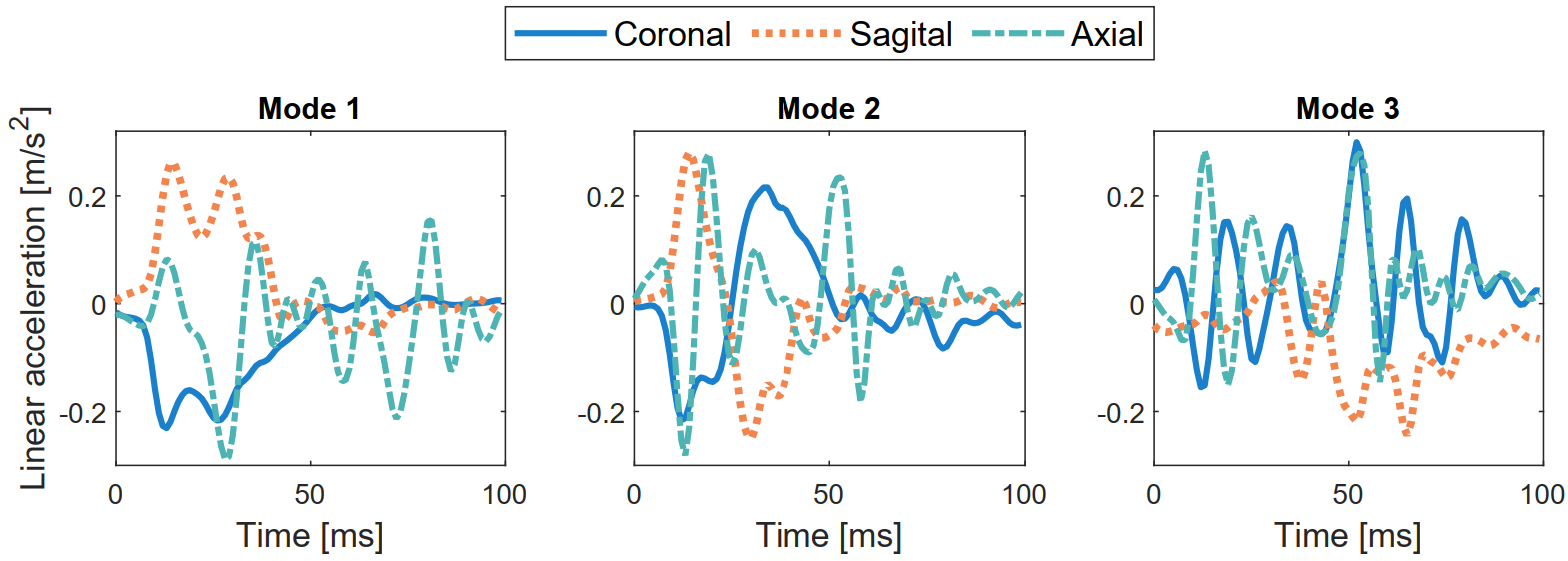}
    \caption{The three most relevant temporal modes for linear acceleration for the entire dataset.}
    \label{fig:tmodesa}
\end{figure}

\begin{figure}[h]
    \centering

    \includegraphics[scale=0.5]{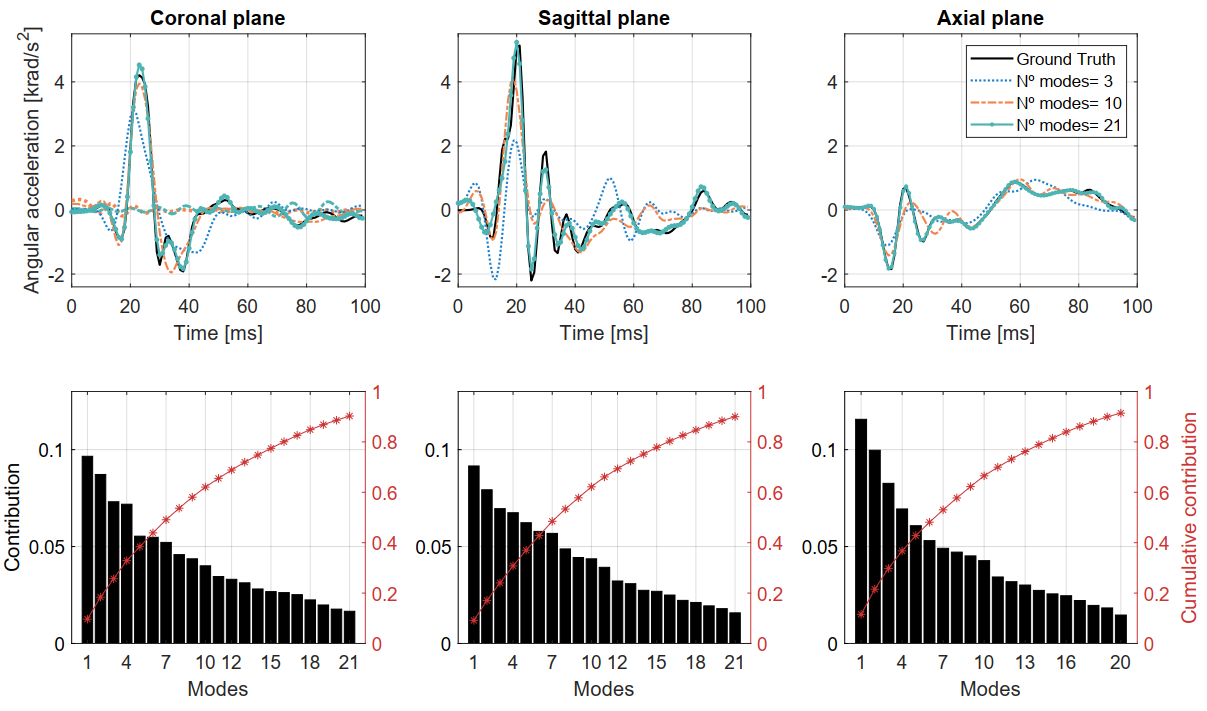}
    \caption{{\it Top:} Representative reconstruction of the angular acceleration data for a single head impact. {\it Bottom:} Individual and cumulative contribution of the temporal modes derived through PCA for angular velocity in each anatomical direction, constituting over 90\% of the ground truth measurements.}
    \label{fig:podalpha} 
\end{figure}

\begin{figure}[h]
    \centering
    \includegraphics[scale=0.4]{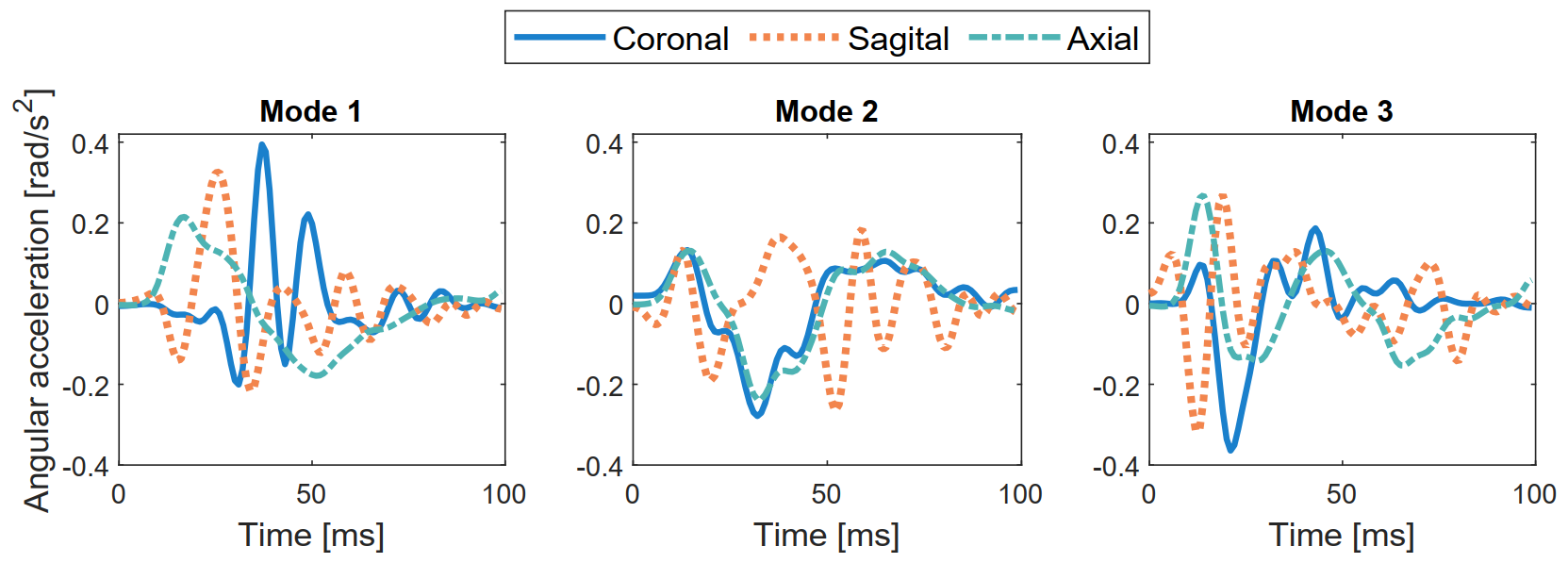}
    \caption{The three most relevant temporal modes for angular acceleration for the entire dataset.}
    \label{fig:tmodesalpha}
\end{figure}

\begin{table}[h]
\centering
\caption{$p$-values for the different injury metrics obtained by Friedman test.}
\begin{tabular}{|c|c|c|c|}
\hline
Parameter & PCA & Tri. & H.S \\ \hline
$HIC_{15}$ & $0.7919$ & $<0.001$ & $<0.001$ \\ \hline
$RIC_{36}$ & $0.1478$ & $<0.001$ & $<0.001$ \\ \hline
$BrIC$ & $0.636$ & $0.380$ & $0.311$ \\ \hline
Coronal BAM & $0.955$ & $<0.001$ & $<0.001$ \\ \hline
Sagittal BAM & $0.917$ & $0.007$ & $0.001$ \\ \hline
Axial BAM & $0.846$ & $<0.001$ & $<0.001$ \\ \hline
MPS WB & $0.731$ & $0.001$ & $<0.001$ \\ \hline
MPS CC & $0.780$ & $<0.001$ & $<0.001$ \\ \hline
FS CC & $0.807$ & $<0.001$ & $<0.001$ \\ \hline
\end{tabular}
\label{Pvaluestable}
\end{table}

\begin{figure}[h]
    \centering
    \includegraphics[width=\textwidth]{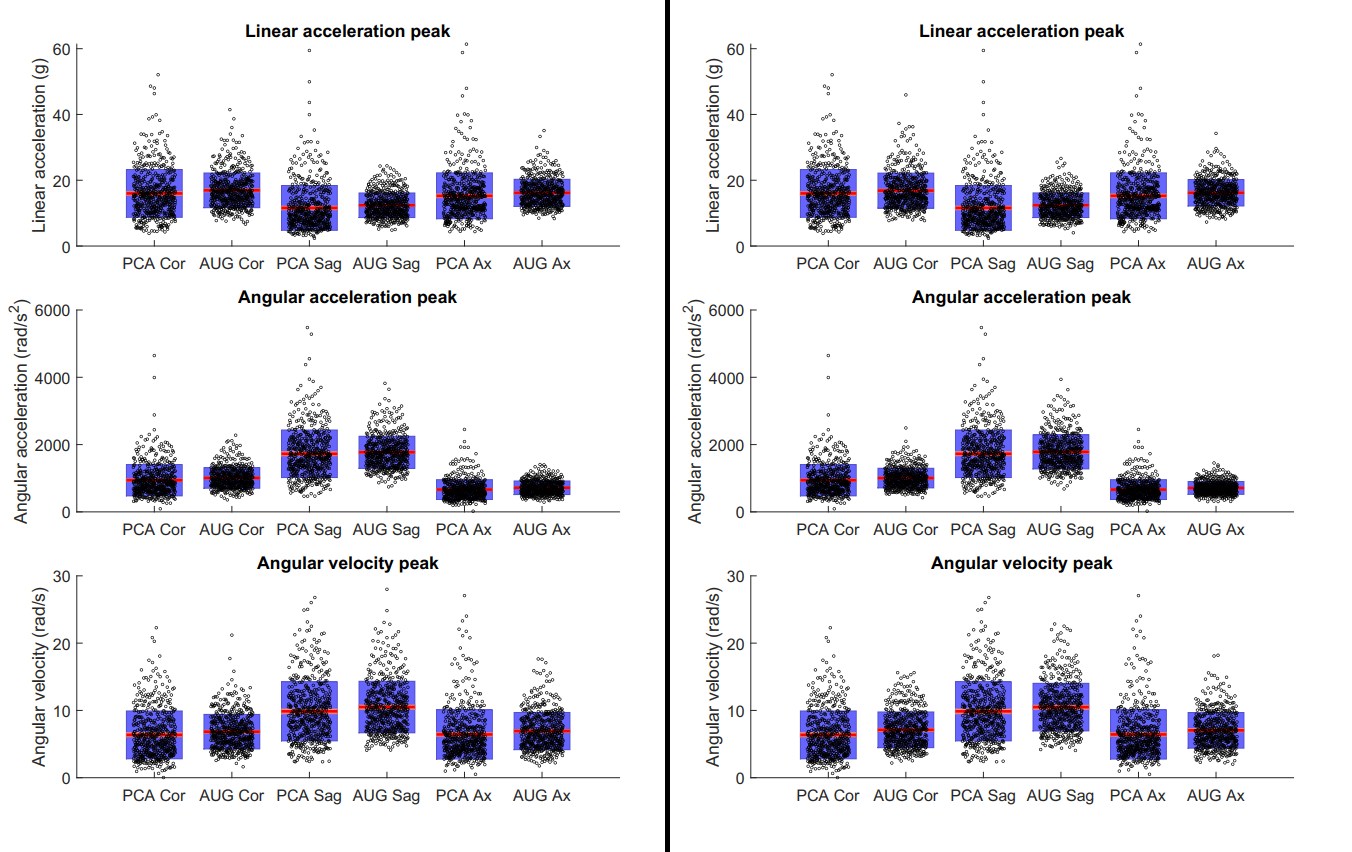}
    \caption{Peak distribution comparison between augmented data (two tests) and PCA original data.}
    \label{fig:peaks}
\end{figure} 

\begin{figure}[h]
    \centering
    \includegraphics[width=\textwidth]{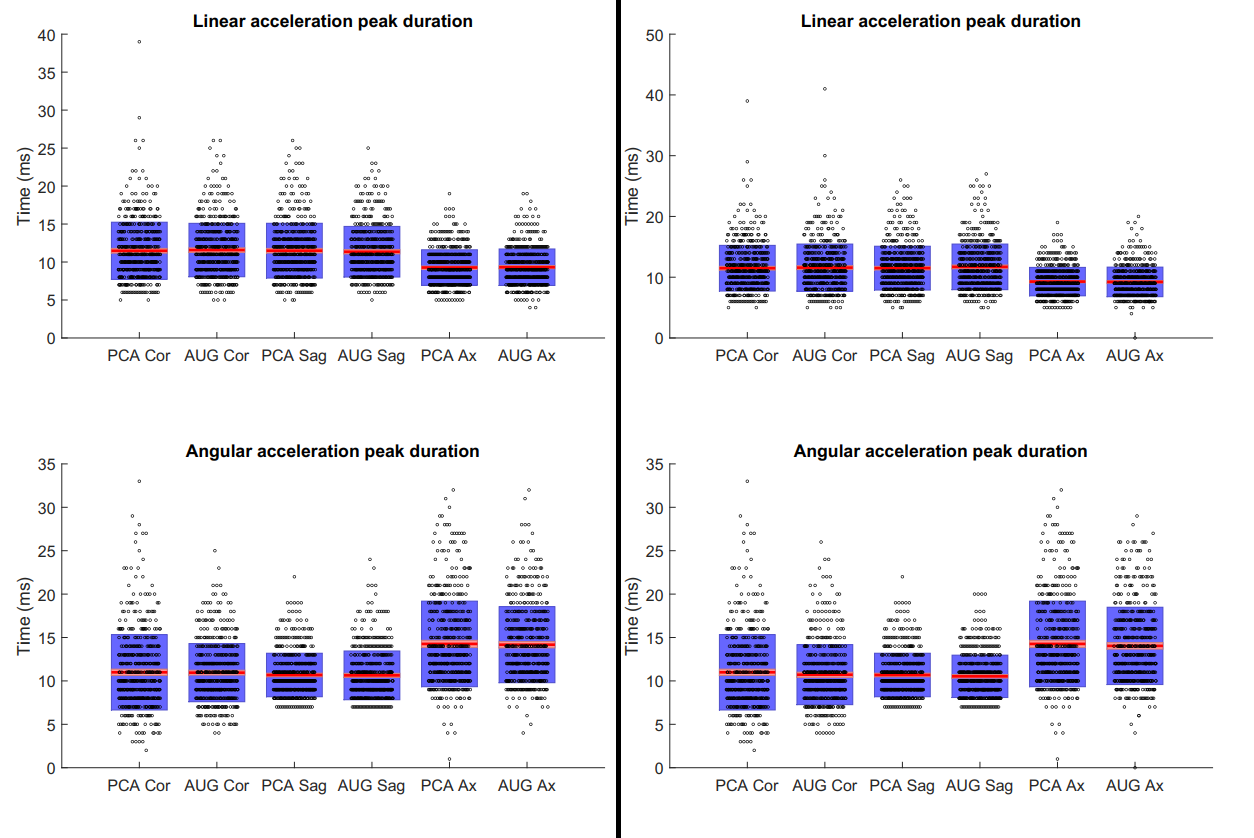}
    \caption{Peak duration distribution comparison between augmented data (two tests) and PCA original data.}
    \label{fig:duration}
\end{figure} 

\begin{figure}[h]
    \centering
    \includegraphics[width=\textwidth]{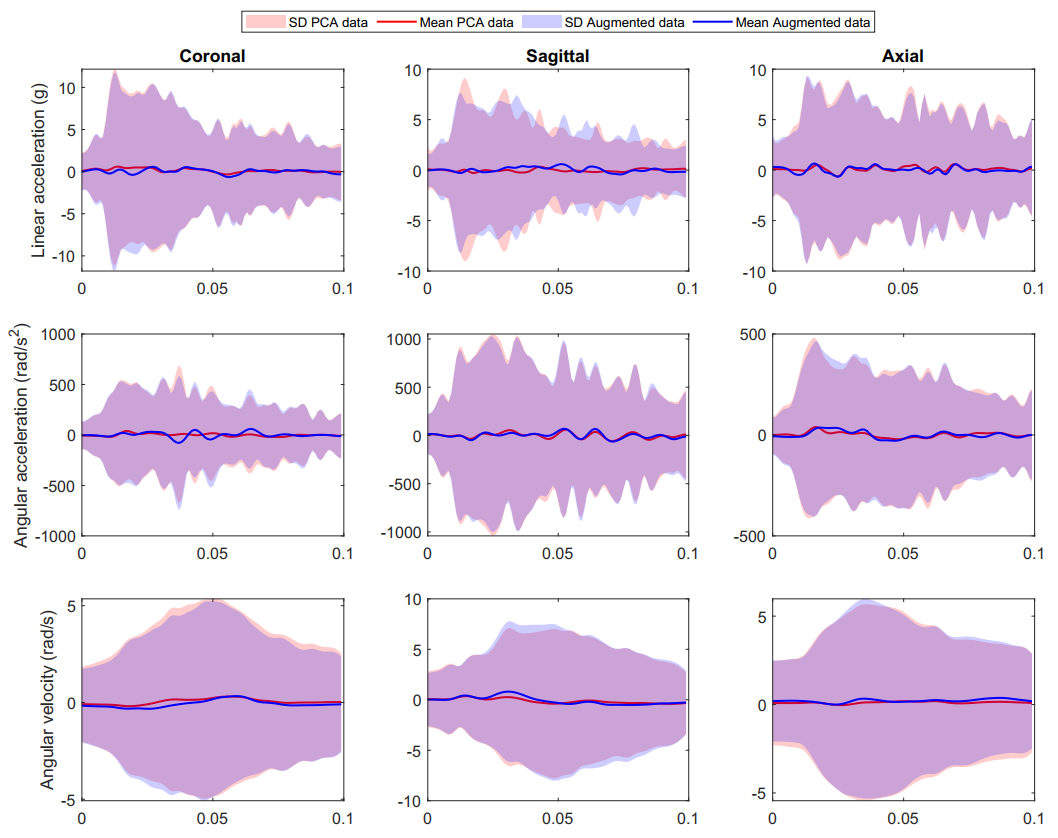}
    \caption{Statistical parameters per time point for PCA and an augmented dataset.}
    \label{fig:statspertime}
\end{figure}

\begin{figure}[h]
    \centering
    \includegraphics[width=\textwidth]{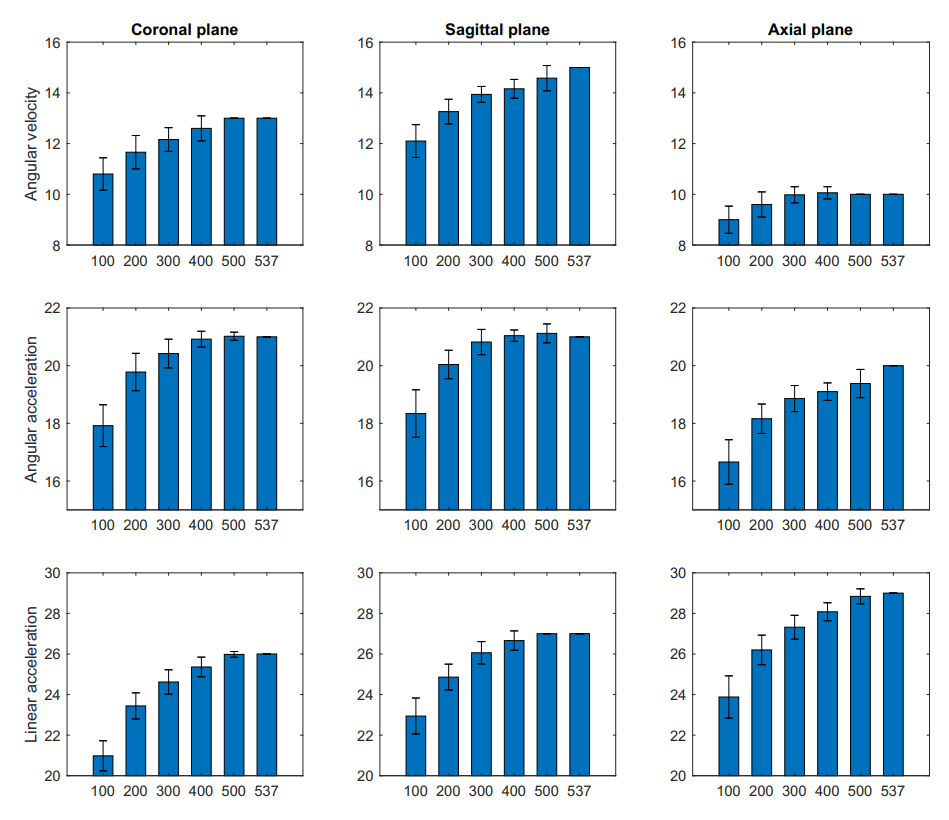}
    \caption{Required number of modes to satisfy Equation \ref{eq:svcriteria} for random subsets of different size, showing convergence to the values obtained with the entire dataset (537).}
    \label{fig:Convergence}
\end{figure}

\end{document}